\documentclass[aps,showpacs,superscriptaddress,preprintnumbers,amsmath,twocolumn,tightenlines]{revtex4}

\usepackage{graphicx,epstopdf,amsmath,amsfonts}
\usepackage{amsmath}
\usepackage{dcolumn}
\usepackage{placeins}
\usepackage{expdlist}

\begin{document}

\bibliographystyle{apsrev}

\title {How do you know if you ran through a wall? }
\author{M. Pospelov}
\affiliation{Department of Physics and Astronomy, University of Victoria,
     Victoria, BC V8P 1A1, Canada}
\affiliation{Perimeter Institute for Theoretical Physics, Waterloo, ON N2J 2W9, Canada}

\author{S. Pustelny}
 \email{pustelny@uj.edu.pl}
 \affiliation{Institute of Physics, Jagiellonian University, Reymonta 4, 30-059 Krak\'ow, Poland}
 \affiliation{Department of Physics, University of California at Berkeley, Berkeley, California 94720-7300}

\author{M.~P.~Ledbetter}
 \affiliation{Department of Physics, University of California at Berkeley, Berkeley, California 94720-7300}

\author{D. F. Jackson Kimball}
 \affiliation{Department of Physics, California State University -� East Bay, Hayward, California 94542-3084, USA}

\author{W. Gawlik}
 \affiliation{Institute of Physics, Jagiellonian University, Reymonta 4, 30-059 Krak\'ow, Poland}

\author{D.~Budker}
 \affiliation{Department of Physics, University of California at Berkeley, Berkeley, California 94720-7300}
 \affiliation{Nuclear Science Division, Lawrence Berkeley National Laboratory, Berkeley, California 94720}

\begin{abstract}

Stable topological defects of light (pseudo)scalar fields can contribute to the Universe's dark energy and dark matter. Currently the combination of gravitational and cosmological constraints provides the best limits on such a possibility. We take an example of domain walls generated by an axion-like field with a coupling to the spins of standard-model particles, and show that if the galactic environment contains a network of such walls, terrestrial experiments aimed at detection of wall-crossing events are realistic. In particular, a geographically separated but time-synchronized network of sensitive atomic magnetometers can detect a wall crossing and probe a range of model parameters currently unconstrained by astrophysical observations and gravitational experiments.

\end{abstract}


\maketitle

{\em Introduction.} Despite a remarkable success of the Standard Model in describing all phenomena in particle physics,
the cosmology presents a formidable puzzle, with dark energy and dark matter - two substances of unknown
origin - comprising about 75\% and 20\% of the Universe's energy budget. Last decades have seen a dramatic expansion of
all experimental programs aimed at clarifying the nature of dark matter (DM) and dark energy (DE). While many widely ranging
theories of dark matter exist, most of the experimental efforts go into searches of dark matter of some particle
physics variety, producing upper limits on the DM-atom interaction strength. Tests of DE models occur on cosmological scale, showing so far its consistency with the cosmological constant.

The purpose of this Letter is to show that a new class of objects, stable topological defects (such as monopoles, cosmological strings or domain walls), that will contribute both to the DM and DE, can be searched for and studied with the global
network of synchronized atomic magnetometers. To be more specific, we consider an example of a domain wall network built from the
axion-like fields. Our focus on axion-like fields and the pseudoscalar interaction of these fields with matter, is motivated by the theoretical considerations of ``technical naturalness", that allow preserving the lightness of the pseudoscalar fields despite a significant strength of interaction with matter. Observable effects of light pseudoscalar particles can vary considerably, depending on their mass $m_a$. We refer the reader to a sample of  literature on the subject,  covering  a wide range of $m_a$ from $10^{-33}$ to $10^5$ eV \cite{sample}.

Scalar-field potentials with some degree of discrete symmetries admit domain-wall-type solutions interpolating between domains of different energy-degenerate vacua \cite{aDW}. In these models, initial random distribution of the scalar field in the early Universe leads to the formation of domain-wall networks as the Universe expands and cools. For QCD-type axions, if  stable, such domain walls could lead to disastrous consequences in cosmology by storing too much energy \cite{aDW}. For an arbitrary scalar field, where parameters of the potential are chosen by hand, the ``disaster" can be turned into an advantage. Indeed, over the years there were several suggestions how a network of domain walls could be a viable candidate for DM or DE \cite{Press,mura}.

Herein, we revisit a subset of these ideas from a pragmatic point of view. We would like to address the following questions: (1) if a network of domain walls formed from axion-like fields exists in our galaxy, what are the chances for an encounter between the Solar system and a pseudoscalar domain wall? and (2) how could the event of a domain-wall crossing the Earth be experimentally determined? Given gravitational constraints on the average energy density of such walls and constraints on the coupling of axion-like fields to matter \cite{spin,Raffelt}, it is not obvious that the allowed parameter range would enable detection. Yet we show in this Letter that there is a realistic chance for the detection of the domain walls, even when the gravitational and astrophysical constraints are taken into account. This goal can be achieved with correlated measurements from a network of optical magnetometers with sensitivities exceeding  $1~{\rm pT}/\sqrt{\rm Hz}$, placed in geographically distinct locations and synchronized using the global positioning system (GPS).

{\em Physics of light pseudoscalar domain walls.}
We start by considering the Lagrangian of a complex scalar field $\phi$, invariant under $Z_N$-symmetry, $\phi\to\exp(i2\pi k/N)\phi$, where $k$ is an integer. We choose the potential in such a way that it has $N$ distinct minima
\begin{eqnarray}
\label{ZN}
{\cal L}_\phi =  |\partial_\mu \phi|^2 - V(\phi) ;~~ V(\phi) = \frac{\lambda}{S_0^{2N-4}}
\left|   2^{N/2}   \phi^N - S_0^N\right|^2\!\!,
\end{eqnarray}
where $S_0$ has dimension of energy and $\lambda$ is dimensionless.

Choosing $\phi = 2^{-1/2}S\exp(i a/S_0)$ to parameterize the scalar field, we find that the potential $V(\phi)$ is minimized for the following values of $S$ and $a$,
\begin{equation}
S=S_0;~~ a=S_0\times \left\{ 0; ~ \frac{2\pi}{N};~\frac{4\pi}{N};...~\frac{2\pi(N-1)}{N}
\right\}.
\label{minima}
\end{equation}
Freezing the Higgs mode to its minimum, $S=S_0$, produces the effective Lagrangian for the $a$ field,
\begin{equation}
\label{La}
{\cal L}_a = \frac12 (\partial_\mu a)^2 - V_0 \sin^2\left(\frac{Na}{2S_0} \right),
\end{equation}
with $V_0 = 4\lambda S_0^4$.
The spatial field configuration $a({\bf r})$ interpolating between two adjacent minima represents a domain-wall solution. A network of intersecting domain walls is possible for $N\geq 3$. The solution for a domain wall along the $xy$-plane that interpolates between $a=0$ and $2\pi S_0/N$ neighboring vacua with the center of the wall at $z=0$ takes the following form,
\begin{equation}
\label{wall}
a(z)  = \frac{4S_0}{N}\times \arctan\left[\exp(m_az)\right]; ~~ \frac{da}{dz} = \frac{2S_0m_a}{N\cosh(m_az)}.
\end{equation}
The characteristic thickness of the wall $d$ is determined by the mass $m_a$ of a small excitation of $a$ around any minimum, $d\sim 2/m_a$. The mass $m_a$ can be expressed in terms of the original parameters of the potential, $m_a = N S_0^{-1}(V_0/2)^{1/2}= (2\lambda)^{1/2}NS_0$. Owing to the fact that $V(\phi)$ can have many different realizations other than (\ref{ZN}),
we shall use solution (\ref{wall}) as an example, rather than a generic  domain-wall profile for $N\geq3$. The important parameters are  the gradient of the field inside the wall, $m_aS_0/N$, and $m_a$, which determines the wall thickness.

{\em Gravitational and astrophysical constraints.}
From the macroscopic point of view at distance scales much larger than $d$, the wall can be characterized by its mass per area,
referred to as tension,
\begin{equation}
\sigma = \frac{\rm Mass}{\rm Area} = \int dz \left| \frac{da}{dz}\right|^2= \frac{8S_0^2 m_a}{N^2}.
\end{equation}
The network of domain walls will have an additional distance-scale parameter $L$, an average distance between walls, or a characteristic size of a domain. This parameter is impossible to calculate without making further assumptions about the mechanisms
of wall formation and evolution. We treat it as a free variable and constrain the maximum energy density of the domain walls, $\rho_{\rm DW} \sim \sigma/L$ in the neighborhood of the Solar System  by the dark-matter energy density, $\rho_{\rm DM} \simeq 0.4~ {\rm GeV/cm}^{3}$,
\begin{equation}
\label{rho}
\rho_{\rm DW} \leq\rho_{\rm DM}  \Longrightarrow \frac{S_0}{N} \leq 0.4{\rm~TeV} \times
\left[\frac{ L}{\rm 10^{-2}~ly } \times  \frac{\rm neV}{m_a} \right]^{1/2}.
\end{equation}
This constraint implies some flexible evolution of the domain-wall network and the possibility for them to build up their mass inside galaxies. We consider such the constraint as the most conservative, i.e. giving the most relaxed bound on $\rho_{\rm DW}$. If the network of domain walls is ``stiff" and its density inside galaxies is not enhanced relative to an average cosmological value, then a stronger constraint  can be derived by requiring that domain walls provide  a (sub)dominant contribution to the dark-energy density, $\rho_{\rm DW} \leq\rho_{\rm DE}$, where $\rho_{\rm DE} \simeq 0.4\times 10^{-5}$~GeV/cm$^3$ \cite{Peebles2003}. In that case the constraint on $S_0/N$ is strengthened by $\sim 300$. Our choice of the normalization for $L$ and $m_a$ in (\ref{rho}) is suggested by the requirement of having wall crossings within $\sim$10 yr with relative velocity of $v=10^{-3}c$ typical for galactic objects, and having the signal duration  in excess of 1 ms. This choice can be  self-consistent within the cosmological scenario for the formation of the domain-wall network from randomly distributed initial $a_{\rm in}$, assuming that the network is ``frustrated", and exhibits $\rho_{\rm DW} \sim R^{-1}$ scaling, where $R$ is the cosmological scale factor. As a word of caution, we add that the numerical simulations of domain walls in some scalar field theories have shown much faster redshifting of $\rho_{\rm DW}$, and never
achieved the frustrated state \cite{Sousa:2009is}. In light of this, some unorthodox cosmological/astrophysical scenarios for the formation of domain walls may be required.

We consider two types of pseudoscalar coupling of the field $a$ with the axial-vector current of a standard-model fermion,
$J^{\mu} = \bar \psi\gamma_\mu\gamma_5 \psi$,
\begin{eqnarray}
\label{lin}
{\cal L}_{\rm lin} = J^{\mu} \times i\phi
\overleftrightarrow{\partial}_\mu \phi^* \times \frac{1 }{S_0f_a } \, \longrightarrow
J^{\mu} \times \frac{\partial_\mu a }{f_a }
\\
{\cal L}_{\rm quad} =J^{\mu}\times\partial_\mu V(\phi) \times
\frac{4S_0^2}{ (f_a'N)^2V_0 }  \,\longrightarrow
J^{\mu} \times \frac{\partial_\mu a^2 }{(f'_a)^2}
\label{quad}
\end{eqnarray}
where the arrows show the reduction of these Lagrangians at the minima of $V(a)$, and $f_i,f_i'$ are free parameters of the model with dimension of energy.  The normalization is chosen in a way to make connection with  axion literature. The derivative nature of these interactions softens problems with ``radiative destabilization" of $m_a$. It is also important that the effective energy parameters normalizing all higher dimensional interactions in (\ref{lin}) and (\ref{quad}) are  assumed to be above the weak scale.
Both ${\cal L}_{\rm lin} $ and ${\cal L}_{\rm quad} $ lead to the interaction of spins ${\bf s}_i$ of atomic constituents and the gradient of the scalar field,
 \begin{eqnarray}
H_{\rm int} &=& \sum_{i=e,n,p}
 2{\bf s}_i\cdot [f_i^{-1}\nabla a +(f'_i)^{-2} \nabla a^2],
\label{Hint}
\end{eqnarray}
For  light scalars of interest, the astrophysical bounds  limit  $|f_{n,p,e}| > 10^{9} $ GeV \cite{Raffelt}, while bounds on quadratic $\partial_\mu a^2$ interactions are {\em significantly weaker},  $f'_i >  10~{\rm TeV}$ \cite{OP}. In what follows we will derive the signal from $f_i$ in (\ref{Hint}), and then generalize it to the $f'_i$ case.

{\em Spin signal during the wall crossing.}
The principles of sensitive atomic magnetometry are, for example, described in Ref.~\cite{Budker2007Optical}. A typical device would use paramagnetic atomic species such as K, Cs, or Rb by themselves or in combination with diamagnetic atoms whose magnetic moments are generated by nuclear spin (e.g., the spin-exchange-relaxation-free [SERF] $^3$He-K magnetometer described in Ref.~\cite{Romalis}). Specializing  (\ref{Hint}) for the case of two atomic species, $^{133}$Cs in the $F=4$ state and  $^3$He in the $F=1/2$ state, we calculate the energy difference $\Delta E$ between the $F_z=F$ and $F_z=-F$ states in the middle of the wall,
\begin{eqnarray}
\nonumber
H_{\rm int} = \frac{{\bf F\cdot \nabla} a}{F f_{\rm eff}};~f_{\rm eff}^{-1}({\rm Cs}) = \frac{1}{f_e}-\frac{7}{9f_p};
~ f_{\rm eff}^{-1}({\rm He})=\frac{1}{f_n};\\
\Delta E=\frac{ 4S_0m_a}{Nf_{\rm eff}}\simeq 10^{-15}\, {\rm eV} \!\times\! \frac{m_a}{\rm neV}\!\times\!
\frac{10^9\,\rm GeV}
{f_{\rm eff}}\!\times\! \frac{S_0/N}{0.4\, \rm TeV},\nonumber\\
 \label{DeltaE}
\end{eqnarray}
In these formulae we assumed that the nuclear spin  is mostly due to unpaired neutron ($^3$He) or $g_{7/2}$ valence proton ($^{133}$Cs), and one can readily observe complementary sensitivity to $f_i$ in two cases. We can express these results in terms of the equivalent  ``magnetic field" inside the wall using $\mu{\bf B}_{\rm eff} {\bf F}/F =\nabla a {\bf F}/(Ff_{\rm eff})$ identification, where $\mu$ is the nuclear magnetic moment. The magnitude of $B_{\rm eff}$ (direction is impossible to predict) is given by
\begin{eqnarray}
B_{\rm eff}^{\rm max}\simeq  \frac{m_a}{\rm neV}\times \frac{10^9\,\rm GeV}{f_{\rm eff}}\times \frac{S_0/N}{0.4\, \rm TeV}
\times\left\{\begin{array}{c}
10^{-11}\, {\rm T} ~({\rm Cs})\\
-10^{-8}\, {\rm T} ~({\rm He})
\end{array}\right.\!\!\!,\nonumber\\
 \label{eq:Beff}
\end{eqnarray}
and the larger equivalent field strength for $^3$He originates from its smaller magnetic moment. The couplings and wall parameters in Eq.~(\ref{eq:Beff}) are normalized to the maximum allowed values from Eq.~(\ref{rho}). The duration of the signal is given by the ratio of wall thickness to the transverse component of the relative Earth-wall velocity,
\begin{equation}
\Delta t \simeq \frac{d}{v_\perp}= \frac{2}{m_a v_\perp} = 1.3 \,{\rm ms}\times \frac{\rm neV}{m_a} \times \frac{10^{-3}}{v_\perp/c}.
\end{equation}
Such crossing time can easily be in excess of the Cs magnetometer response time $t_r$, and we can combine the $B_{\rm eff}^{\rm max}$ and $\Delta t$ into a signal factor ${\cal S} = B_{\rm eff}^{\rm max} (\Delta t)^{1/2}$ to be directly compared to experimental sensitivity,
\begin{eqnarray}
{\cal S}
\simeq \frac{0.4\, \rm pT}{\sqrt{\rm Hz}}\times \frac{10^9\,\rm GeV}{f_{\rm eff}}\times \frac{S_0/N}{0.4\, \rm TeV}\times
\left[\frac{m_a}{\rm neV} \frac{10^{-3}}{v_\perp/c}\right]^{1/2}  \nonumber\\
\leq \frac{0.4\, \rm pT}{\sqrt{\rm Hz}}
\times \frac{10^9\,\rm GeV}{f_{\rm eff}}\times \left[ \frac{ L}{\rm 10^{-2}~ly }\frac{10^{-3}}{v_\perp/c}\right]^{1/2}\!\!\!\!\!,\;\;\;\;\;\;
\label{S}
\end{eqnarray}
where in the inequality we used the gravitational constraint from Eq.~(\ref{rho}). The maximally allowed value for the signal ($\sim {\rm pT/\sqrt{Hz}}$), after taking into account the gravitational and astrophysical constraints, exceeds capabilities of modern magnetometers that can deliver fT/$\sqrt{\rm Hz}$ sensitivity \cite{Budker2007Optical}. For the $^3$He-K SERF magnetometer, the more appropriate figure of merit would be the tipping angle of the helium spin after the wall crossing, assuming that the typical crossing time is below the dynamical response time. Taking the spins to be oriented parallel to the wall, we calculate this angle to be
\begin{equation}
\Delta \theta = \frac{4 \pi S_0}{v_\perp Nf_{\rm eff}}  \simeq 5\times 10^{-3}\,{\rm rad}\times \frac{10^9\,\rm GeV}{f_{\rm eff}}\times
\frac{10^{-3}}{v_\perp/c}\times \frac{S_0/N}{0.4\, \rm TeV}.
\label{angle}
\end{equation}
This could be far in excess of 10-nrad tipping angles that can be experimentally detected \cite{Kornack2005Nuclear}. Thus, both types of magnetometers offer ample opportunities for a realistic detection of the wall-crossing events. So far we have used the galactic constraints (\ref{rho}), $\rho_{\rm DW} \leq\rho_{\rm DM} $. It is noteworthy that  even if the energy density of walls
in the galaxy does not exceed cosmological dark-energy density, i.e. $\rho_{\rm DW} \leq\rho_{\rm DE} $, the expected signal can reach $\Delta \theta \sim 10^{-5}$~rad and ${\cal S} \sim {\rm fT/\sqrt{Hz}}$, which is still a realistic signal for detection with the best magnetometers. It is remarkable that a possible domain-wall component of DE can, in principle, be detected in the laboratory.

Going over to $f'$ couplings, we notice that the structure of the signal is different: $B_{\rm eff}^{\rm max}$ now changes sign, vanishing in the middle of the wall. Taking  $B_{\rm eff}^{\rm max}$ at $a= S_0\pi/(2N)$ inside the wall, and skipping intermediate states in a similar derivation, our sensitivity formulae (\ref{eq:Beff}) and (\ref{S}) are modified according to the following substitution,
\begin{eqnarray}
\frac{10^9~{\rm GeV}}{ f_{\rm eff} }\longrightarrow 0.6\times 10^4 \times  \left(\frac{10~{\rm TeV}}{f'_{\rm eff}}\right)^2 \times \frac{S_0/N}{0.4~{\rm TeV}},
\end{eqnarray}
where again $f'$ is normalized on its minimum allowed value. One can observe a dramatic increase in the possible signal due to a much weaker astrophysical constraints on ${\cal L}_{\rm quad}$. In Fig. 1, we plot the experimental accessible parameter space in terms of characteristic time between wall crossing events, $T=L/(10^{-3} c)$, and strength of the coupling constants, $f$ and $f'$, fixing $m_a = 10^{-9}~eV,v_\perp/c=10^{-3}$ for concreteness, and saturating either DM or DE density constraints. We assume that the magnetometer sensitivity is ${\cal S} = {\rm fT/\sqrt{Hz}}$. The light(dark) shaded areas indicate the coupling range that  can be realistically probed with the magnetometer network when DM(DE) constraints are saturated, by imposing all constraints and additionally requiring $T<10$ yr. One can see that the large part of the parameter space is accessible, and for the case of ${\cal L}_{\rm quad}$ even the DE constraint can allow for a detectable signal with $T<1$~yr.
\begin{figure}[hbtp]
\centerline{\includegraphics[trim=0.85cm 0.75cm 1.5cm 2cm, width=\columnwidth]{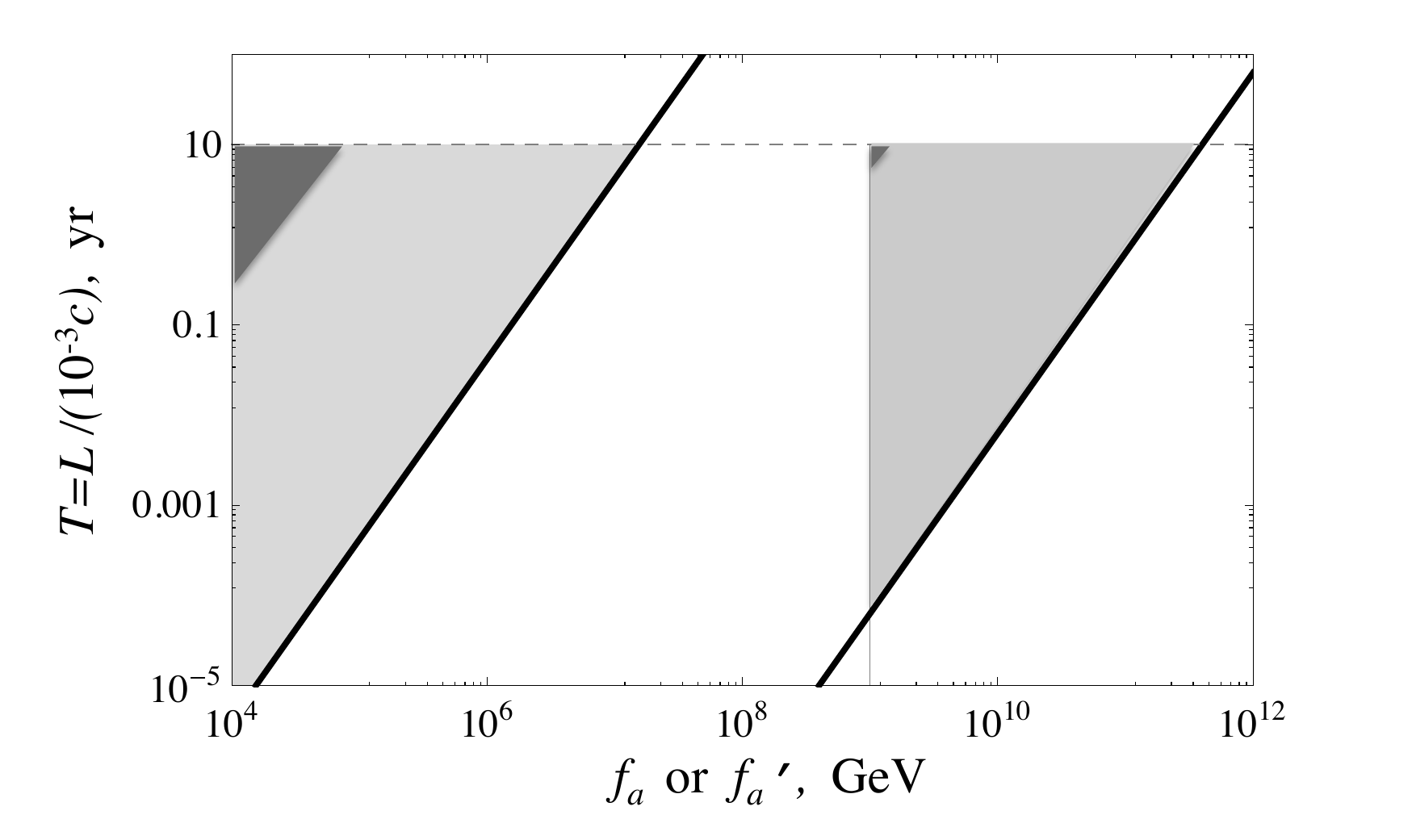}}
 \caption{\footnotesize Parameter space open for detection of the wall crossing, $T/(10^{-3} c)$ in yr vs $f(f')$ in GeV. Shaded triangle on the left correspond to ${\cal L}_{\rm quad}$ case, and on the right to  ${\cal L}_{\rm lin}$. }
\end{figure}

{\em Network of synchronized magnetometers.} While a single magnetometer is sensitive enough to detect a domain-wall crossing, due to the rarity of such events it would be exceedingly difficult to confidently distinguish a signal from false positives induced by occasional abrupt changes of magnetometer-operation conditions, e.g., magnetic-field spikes, laser-light-mode jumps, etc. A global network of synchronized optical magnetometers is an attractive tool to search for galactic/cosmological domain walls, as it would allow for efficient vetoes of false domain-wall crossing events.

Ideally, one would require $n\geq 5$ magnetometer stations in such a network. The difference in timing $t_i$ of a putative signal
is related to the transverse velocity and the unit normal vector to the wall, ${\bf n}$, $t_i-t_j  = {\bf L}_{ij}\cdot {\bf n} v_\perp^{-1}$, where ${\bf L}_{ij}$ are the three-vectors of the relative positions of magnetometers $i$ and $j$. Four stations are required to specify magnetometer-defined 3D system of coordinates, and three time intervals between four $t_i$ will enable to unambiguously determine the three-vector ${\bf n}v_\perp^{-1}$. This makes the predictions for the timing of the event at the fifth station, $t_5$, which can be used as a tool for rejecting accidental backgrounds. Consider a network of similar magnetometers with fast response time separated by distances of $O(300 ~\rm km)$  operating during a long period  ${\cal T}\sim {\rm yr}$.
Suppose that $\tau$ is an average time between accidental spikes in the background above certain value $B_{\rm eff}^0$ that cannot be distinguished from the signal. Then the probability of having four events in four different stations within time intervals corresponding to the typical wall travel time from station to station, $t_{\rm trav} \sim l/v \sim {\rm s}$, is $P_{1234}\sim{\cal  T }t_{\rm trav} ^3 \tau^{-4}$, where we take ${\cal T}\gg \tau \gg t_{\rm trav}$. To have this probability below one, one should achieve $\tau > 100\,{\rm s}$.  If indeed four accidental background spikes lead to false signals in four stations within $t_{\rm trav}$, the domain wall interpretation will predict the event in the fifth station within a narrow window of the wall crossing $\Delta t\sim {\rm ms}$, and the probability of this to happen due to accidental background is $P_{12345} \sim (\Delta t/\tau)P_{1234}$, or less than $10^{-5}$ for $\tau \sim 100\,{\rm s}$. Increasing the number of stations will enable to search
for weaker signal $B_{\rm eff}^0$, and tolerate shorter $\tau$ \cite{Abbott2004}.

Recently we set up a prototype for the magnetometer network consisting of two magnetometers operated in magnetically shielded environments located in Krak\'ow, Poland and Berkeley, USA (a separation distance of about 9000~km). One of the magnetometers (Krak\'ow) is based on nonlinear magneto-optical rotation \cite{Pustelny2008Magnetometry}, while the other magnetometer (Berkeley) is a SERF device \cite{Ledbetter2008Spin}.
\begin{figure}[hbtp]
\centerline{\includegraphics[trim=0.85cm 0.5cm 2cm 0cm, width=6cm]{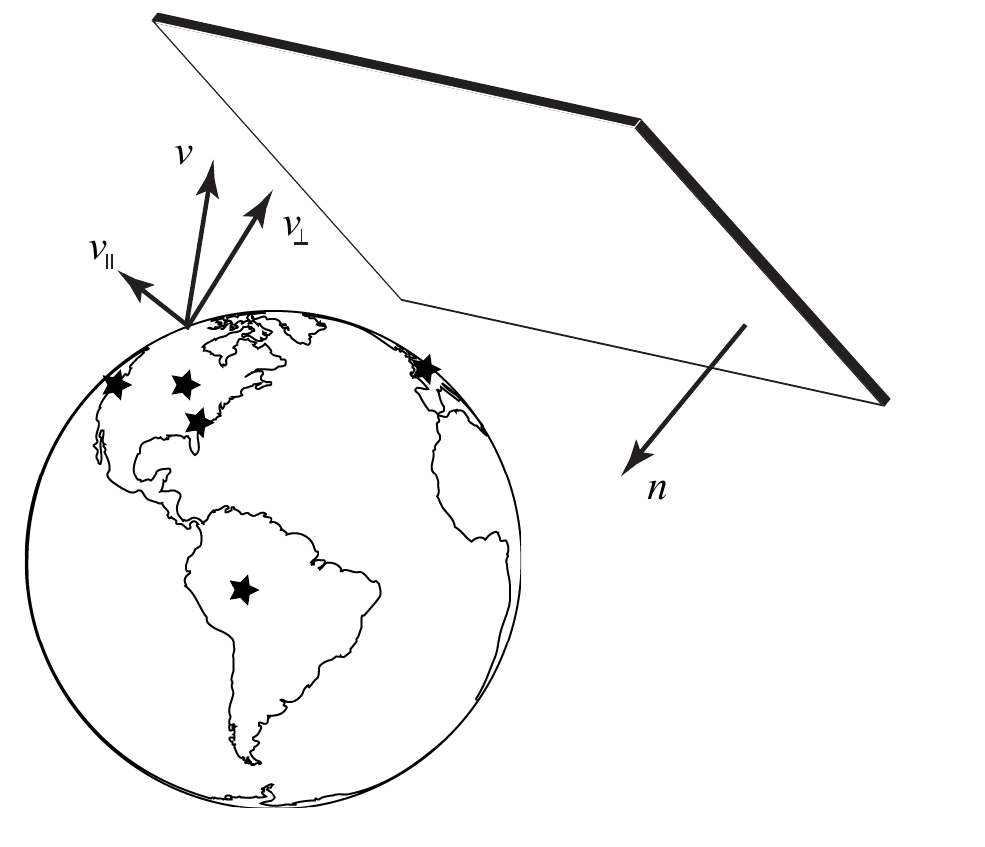}}
 \caption{\footnotesize Schematic of Global Network of Optical Magnetometers for Exotic physics (GNOME) planned to be used for detection of the domain-wall crossing. The wall-crossing events recorded with four magnetometers at $t_i$ allow determination of the normal velocity of the wall $v_\perp$. The remaining magnetometer(s) will be used to verified the measurements by predicting the time of the events in the locations (see text).}
\label{bang}
\end{figure}
The magnetometers achieved comparable sensitivities of 10~fT/Hz$^{1/2}$, which can be further improved upon optimization. The expected parameters of the signal, $\Delta t \sim$ 1 ms and the minimum time-separation between the events $\Delta t_{\rm trav} \sim 30\,{\rm s}$, can be precisely determined using a GPS time source (for more details see Ref.~\cite{Pustelny2012Global}). We have recently performed proof-of-principle experiments \cite{Pustelny2012Global} demonstrating the ability to correlate the signals of two magnetometers. In particular, we demonstrated significant reduction of noise and rejection of false-positive events present in magnetometer signals. The measurements proved the feasibility of correlated magnetic-field measurements opening avenues for further investigations involving more magnetometers.

{\em Summary}. We have shown that a network of modern magnetometers offers a realistic chance for detecting the event of an axion-type domain-wall crossing and can probe parts of the parameter space where such walls can contribute significantly to
dark matter/dark energy.

The authors are grateful to N. Afshordi, A. Arvanitaki, A. Derevianko, J. Brown, S. Carroll, M. Kozlov, V. Flambaum, M. Kamionkowski, and M. Hohensee for discussions. This work was supported in part by the NSF and Polish TEAM program of the Foundation for Polish Science. SP is the scholar of the Polish Ministry of Science and Higher Education within the Mobility Plus program.

\end{document}